**On a preseismic electric field "strange attractor" like precursor observed short before large earthquakes.**

Thanassoulas[1], C., Klentos[2], V., Verveniotis, G.[3]

1. Retired from the Institute for Geology and Mineral Exploration (IGME), Geophysical Department, Athens, Greece.
   e-mail: thandin@otenet.gr - URL: www.earthquakeprediction.gr

2. Athens Water Supply & Sewerage Company (EYDAP),
   e-mail: klenvas@mycosmos.gr - URL: www.earthquakeprediction.gr

3. Ass. Director, Physics Teacher at 2nd Senior High School of Pyrgos, Greece.

**Abstract.**

The Earth's electric field, recorded by two different and distant monitoring sites located in Greece, is analyzed in terms of monochromatic signals (**T=24h**). The analysis is based on the mapping procedures borrowed from the study of non-linear dynamic systems and chaos. The specific analysis applied on the electric field recorded for some weeks before the occurrence of 3 large EQs which occurred in Greece shows some very interesting results. Long before the EQ occurrence the phase map generated by the uncorrelated each other signals recorded by a pair of distant monitoring sites is characterized by pairs of hyperbolas of random azimuth asymptotes. Some days before the occurrence of the earthquake the recorded signals are strongly correlated and therefore the phase map is characterized by ellipses. A couple of days before the EQ occurrence the electric signals become again uncorrelated due to the fact that their generating mechanism (large scale piezoelectricity) approaches its final stage and therefore hyperbolas characterize again the phase map. When this stage is observed the pending EQ shortly (in terms of days) takes place. Three detailed examples from large Greek EQs are presented towards the validation of this mechanism.

**1. Introduction.**

Although in the recent 2-3 decades of years, the number of the research papers, which deal with the relation of the Earth currents to earthquakes, has increased, actually, this knowledge is not a new one. A deep research in the scientific literature has revealed that similar observations have been made, in a simpler form than nowadays, since 1692. The earliest paper that was traced is that of Milne (1890), where a large number of such events were reported.

Research papers which refer to earthquake precursory electric signals were presented by Terada (1931), Fedotov et al. (1970), Sobolev (1975), Sobolev et al. (1972), Varotsos et al. (1981, a), Thanassoulas (1982), Nayak et al. (1982), Varotsos et al. (1982), Ralchovsky and Komarov (1988), Meyer and Pirjola (1986), Miyakoshi (1986), Thanassoulas and Tselentis (1986), Meyer and Ponomarev (1987), Thanassoulas and Tselentis (1993), Ifantis et al. (1993), Tselentis and Ifantis (1996), Enomoto et al., (1997), Thanassoulas and Tsatsaragos (2000), Fujinawa et al. (2000), Zlotnicki et al. (2001), Eftaxias et al. (2001), Karakelian et al. (2002), Nagao et al. (2002), Karakelian et al. (2002a), Honkura et al. (2002), Pham et al. (2002), Thanassoulas and Klentos (2003), Hattori et al. (2004), Ida and Hayakawa et al. (1994, 2006), Varotsos (2006), Thanassoulas (2008, a, b), Thanassoulas et al. (2008).

The literature, earlier mentioned, is a small part of the plethora of papers, which exist in the worldwide, seismological, scientific journals and refer to the generation of earthquake precursory, electrical signals.

On the other hand, a limited number of papers which strongly object the validity of the generation of such electrical, seismic, precursors do exist, too. In contrast to the references, listed above, which are in favor of the generation of seismic, electric precursors, only a very small number was traced against, such as: Gruszow et al. (1996), Variemezis et al. (1997), Bernard et al. (1997), Pinettes et al. (1998), Pham et al. (1999), Variemezis et al. (2000), Pham et al. (2001), Pham et al. (2002).

The majority of afore mentioned papers deals with the time prediction of a large earthquake. On the other hand only a very limited number of them deal with the determination of the epicentral area of the imminent earthquake. Characteristic examples of such papers are these which were presented by Varotsos (1981a), Thanassoulas (1991, 1991a, 2007, 2008), Thanassoulas et al. (2008). The main characteristic of these papers is that the electric field intensity vector, determined in two or more monitoring sites, facilitates through triangulation the determination of the epicentral area. Details of this methodology have been presented by Thanassoulas (2007) in a monograph under the title "Short-term Earthquake Prediction", while more examples from real large earthquakes which occurred in Greece are presented in the URL: www.earthquakeprediction.gr

An exceptional case of the precursory seismic electric signals is the one when a monochromatic (**T=24h**) signal is analyzed. This signal not only suggests that the critical phase before the occurrence of a large EQ has been reached (Thanassoulas and Tselentis 1986, 1993), but at the same time exhibits directional properties (Thanassoulas et al. 2008) which indicate the azimuthal direction of the pending large EQ in respect to the location of the related monitoring site for the Earth's electric field registration.

In this work the monochromatic (**T=24h**) preseismic electric signals recorded by two different distant monitoring sites are analyzed by the use of methods borrowed from "Dynamics" and "Chaotic systems" (Nusse and Yorke, 1998; Korsch et al. 2008). In detail, mapping techniques are used in order to identify any "strange preseismic attractor" which will identify the presence of any preseismic electric signal induced by a future large EQ.

**2. Theoretical analysis.**

The Earth's electric field, registered, by a monitoring site, can be treated in a quite different way than the traditional ones, in an attempt to calculate the regional location of its current source. For this purpose, let us recall the electrical field intensity vector of an oscillating field and its, graphical, spatial presentation of its azimuthal direction as a function of time, presented, in figure **(1).** The major axis of the ellipse (left) indicates the azimuthal direction **(θ)** of the location of the current source in relation to the monitoring site. If a different monitoring site at a different and distant location is used simultaneously, it is obvious that a different ellipse (right) will be produced, but its major axis will point towards the very same current source, too. This is illustrated in the following figure **(2).**

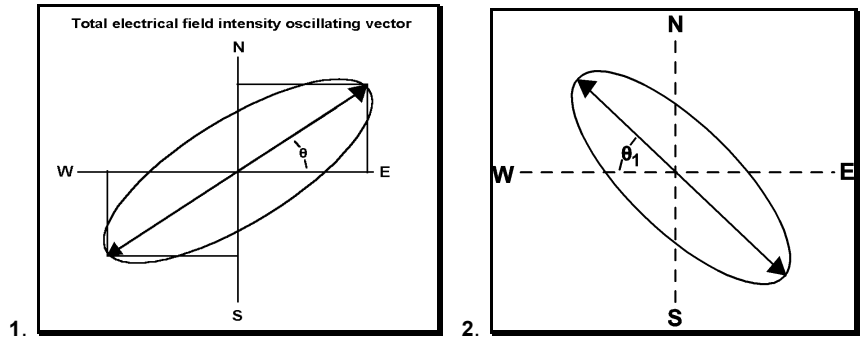

Fig. 1 - 2. **Left**: The oscillating, electric field intensity vector prescribes an ellipse. The angle **(θ),** of its major axis in reference to the **EW** direction, indicates the azimuthal direction of the location of the current source in relation to the location of the monitoring site. **Right**: The angle **(θ1),** of its major axis in reference to the **EW** direction, indicates the azimuthal direction of the location of the same current source, but at a different monitoring site (after Thanassoulas, 2007).

The intersection of these two azimuthal directions **θ** and **θ1,** evidently, indicates the location of the current source that generates the oscillating, electric field. In other words, it indicates the virtual, epicentre area (Thanassoulas, 2007, 2008) of the pending earthquake. The later is illustrated in the following figure **(3).**

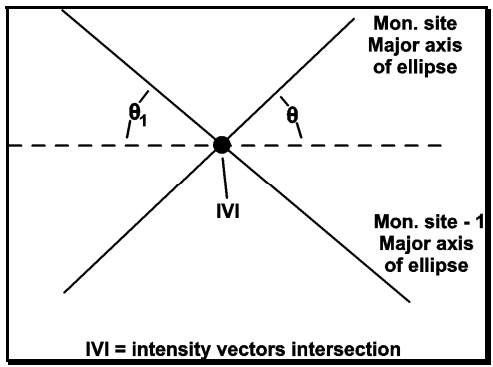

Fig. 3. The two major axis of the ellipses (of azimuthal angles **θ**, **θ1**), observed, at the two monitoring sites, intersect at point **IVI** (**I**ntensity **V**ector **I**ntersection) which is considered as the location of the virtual, epicentre area location.

In a more general case, while the electrical field intensity vector prescribes its specific ellipses, at any time **(t$_i$)** corresponds an azimuthal direction **(θ$_i$).** Therefore, the equation:

$$Y = θ_i * x + b \qquad (1)$$

is time depended. A pictorial presentation of the way equation **(1)** is formed is shown in figure **(4)**.

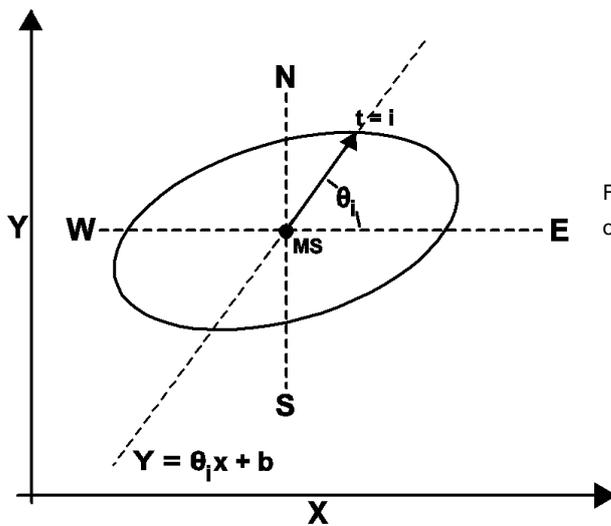

Fig. 4. The linear equation that represents the azimuthal direction of the electric field intensity vector at any time **(t$_i$)** is presented.

A similar equation, but of different **(θ$_i$),** holds for a different monitoring site. For the case of two different monitoring sites the two linear equations form a linear algebraic system. Therefore, it is possible to analyze not only the intersection of the major axis of the corresponding ellipses, but to study, generally, the intersection location as a function of time. The later is presented in algebraic form as follows in the system of equations **(2):**

$$\left. \begin{matrix} Y_i = X_i*\theta_i(t) + b \\ Y_k = X_k*\theta_k(t) + c \end{matrix} \right|_1^n \blacktriangleright [X, Y]_t \qquad (2)$$

The left part of it represents the system of equations which holds for **(n)** data pairs obtained from the two **(i, k)** monitoring sites, while the right part of it represents the space of the algebraic system solutions which corresponds the entire electric field registered data pairs.

The procedure for the implementation, in practice, of the system of equations **(2)** is briefly as follows:

   a.   For each data sample, obtained from both monitoring sites, the azimuthal directions as functions of time are calculated.

   b.   The linear equations as functions of time are formed.

   c.   The location of the **IVI** is determined, from the calculated, azimuthal directions in a phase plane, as a solution of the system of linear equations that forms by each data pair (from the two monitoring sites).

   d.   The obtained **IVI** locations are plotted, on a regional **X – Y** phase map.

The entire procedure in terms of hardware and software is presented in the following figure **(5).**

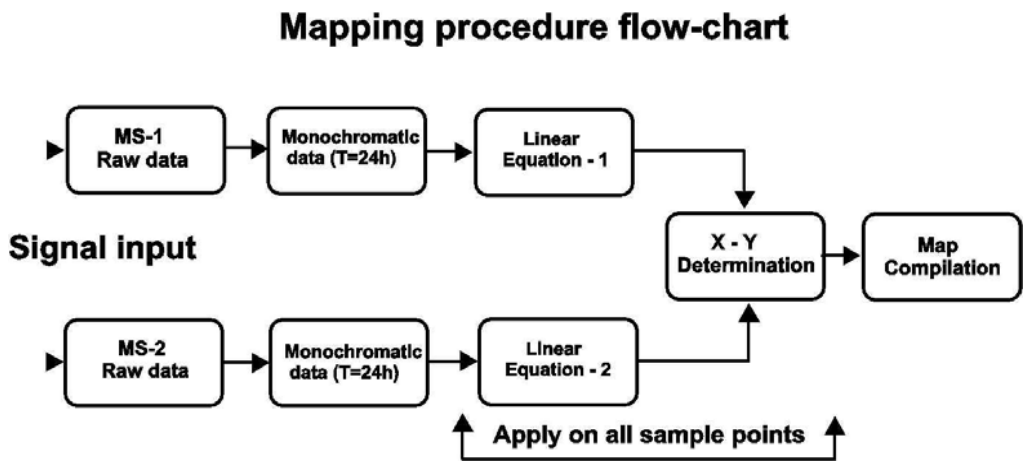

Fig. 5. Flow-chart of the mapping procedure. The raw data which are registered at two different and distant monitoring sites are firstly normalized and band-pass filtered (Thanassoulas, 2007). Next, the linear equations are formed that correspond to each monitoring site and then the linear algebraic system is solved, consecutively for the entire data set. Finally the phase map of the **IVI** is compiled.

The specific methodology was borrowed from the theories which are applied on chaotic and no-linear dynamic systems (Nusse and Yorke, 1998; Korsch, Jodl and Hartmann, 2008).

Two distinct cases may be considered in studying monochromatic oscillating electric fields. The first one is the case when the two signals originate by different sources and therefore are uncorrelated. The second is the case when the two signals have common origin source. The later is the case of the electric field, generated by the focal area of a pending large EQ, which is registered by two different monitoring sites.

**- Uncorrelated electric fields.**

In this case the intersection of the two electric field intensity vectors, as long as the electrical field oscillates, moves in the plane of oscillation. Moreover, at different periods of time, the **IVI** location, in the phase map, will either represents the first major ellipses axis of the first monitoring site or the corresponding one of the second monitoring site. Theoretically the **IVI** will smoothly swing from one azimuthal direction suggested by the first major ellipse axis to the second one. Consequently, it will prescribe, in time, a hyperbolic trace, which asymptotically will coincide with the two major axes of the, observed ellipses by the two different monitoring sites. The later is indicated in the following figure **(6).**

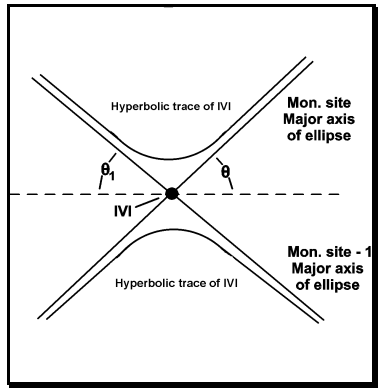

Fig. 6. Hyperbolic traces prescribed by the intersection of the oscillating, electrical field intensity vector, as a function of time. Intersection (of major ellipse axes) of the hyperbolas asymptotes is indicated by **IVI**.

Moreover, since the oscillating signals have been obtained from random electric noise, the major axis calculated at any time for each monitoring site will be randomly time depended as far as it concerns its direction in space. Thus, the pair of hyperbolas prescribed in space will change randomly its orientation in time.

The prevailing direction, and therefore the prescribed hyperbola, at a certain period of time, depends upon the amplitude and phase shift (observed between each other) of each oscillating field determined for the same period of time.

The intersection of the two different random intensity vectors follows random hyperbolic paths, just because there is generally an unknown phase shift and amplitude difference between the oscillating fields, as it is evident from figures **(1)** and **(2).**

**- Correlated electric fields.**

In the case that both signals are generated by the same current source, these are closely interrelated and therefore, instead of hyperbolas, ellipses are generated by the **IVI**. Actually, in real conditions, the theoretical intersection **IVI** of figure **(3)** "explodes" into an ellipse, due to the presence of noisy phase shifts, different electric field amplitudes, which are caused by interfering noisy signals.

Such kind of correlated signals can be generated by the piezoelectric mechanism or any other one that exhibits piezoelectric properties / behaviour. A simplified piezoelectric mechanism is presented in the following figure **(6a)**.

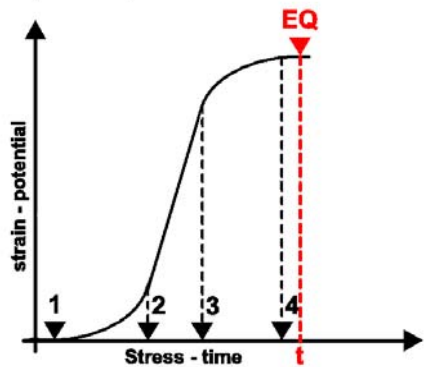

Fig. 6a. A simplified piezoelectric mechanism. Numbers: **1, 2, 3, 4** denote the boundaries of the phases of the rock deformation along time as increasing stress is applied. Increasing non-linear deformation starts at boundary **(1)**, it is followed by an almost linear deformation phase between boundaries **(2)** and **(3)** followed by a non-linear decreasing deformation in between boundaries **(3)** and **(4)**. The rock formation collapses at time **(t)** which denotes the time of occurrence of the EQ.

For the referred deformation phases the corresponding monochromatic **(T=24h)** electric oscillating signal which is triggered by the lithospheric tidal oscillation, develops as follows: non-linear increasing in amplitude electric signal is initiated at boundary **(1)**. Between boundaries **(2)** and **(3)** it acquires its maximum amplitude. Between boundaries **(3)** and **(4)** its amplitude decreases again. Next, in a short time period, the rock formation collapses, in other words the earthquake takes place (Thanassoulas and Tselentis, 1993).

If the piezoelectric mechanism is combined with the mapping procedure of system of equations **(2)** on the data obtained from two different monitoring sites the following results are theoretically expected:

a. Before deformation boundary **(1)** no oscillating signals are expected to occur generated from the focal area. Therefore, the oscillating signals which will be obtained by band-pass filtering of the recorded electrical noise at both monitoring sites will be uncorrelated. Consequently, the mapping procedure will generate typical random in direction hyperbolic traces.
b. There is a possibility that the very same holds as in **(a)** for the period of time between **(1)** and **(2)** deformation boundaries. This could be due to the small amplitude and masking by noise of the registered electric oscillating signals.
c. During the period of time between the deformation boundaries of **(2)** and **(3),** the oscillating electric signal acquires its maximum amplitude and therefore, the registered electric oscillating signals are highly correlated. Consequently, ellipses will be generated by the mapping procedure.
d. During the period of time between the deformation boundaries of **(3)** and **(4)**, the oscillating electric signal decreases drastically in amplitude and therefore for this period of time the mapping procedure will generate typical random hyperbolic traces.
e. Finally, shortly after the ellipses have been replaced again by hyperbolas the earthquake will take place following the typical behaviour of the stress – strain curve which the rock formation follows.

The first example of this methodology was presented by Thanassoulas (2007). It concerns the earthquake which occurred in Greece on 19th April 2007 having an **Ms = 5.4R**. The details of this example are as follows:

The calculated intensity vector traces from recordings of the Earth's electric field, are presented in the following figures **(7, 8, 9).** The used monitoring sites are of **PYR** and **HIO** and the time span of the recording is from 20070401 to 20070404, (dates in yyyymmdd mode).

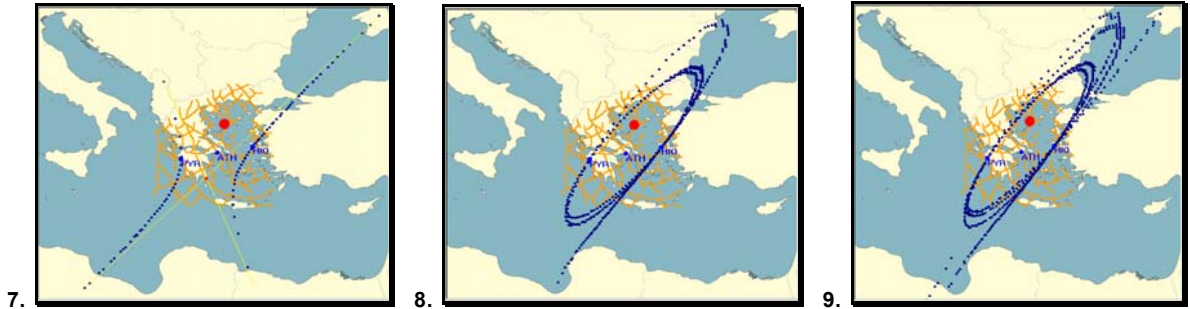

**7.**   **8.**   **9.**

Fig. 7. Typical intensity vector intersections and asymptotes are presented. The red dot indicates the "virtual epicentre" valid for period from 20070401 to 20070404.
Fig. 8 - 9. Ellipses observed, in the period from 20070404 to 20070407 and 20070406 to 20070410. The ellipses suggest the strong interrelationship of the recorded signals. The large, solid, red circle, in all figures, indicates the location of the earthquake (M=5.4R) that occurred on 19th April 2007.

The calculated traces are depicted by blue dots with a sampling interval of ten **(10)** minutes. Thin yellow lines represent the corresponding asymptotes. Their intersection is indicated by a small red dot, while a solid, large red circle indicates the location of the earthquake (**M=5.4R**) that occurred on 19th April 2007.

The previous figure **(7)** transforms gradually, into the figure **(8),** due to the increase of correlation of the recorded signals. The later indicates that a hypothetical preseismic signal "enters" the monitoring network.

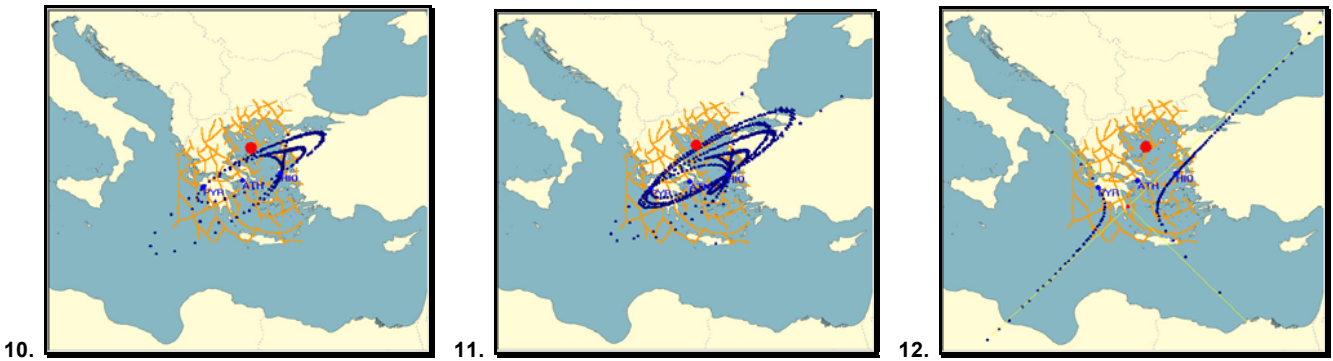

**10.**   **11.**   **12.**

Fig. 10 – 11. Ellipses observed, in the period from 20070412 to 20070415 and 20070414 to 20070417. The ellipses suggest the strong correlation of the recorded signals.
Fig. 12. Hyperbolas observed, in the period from 20070418 to 20070421. It is suggested that the "preseismic signal effect" has disappeared. The large, solid, red circle indicates the location of the earthquake (**M=5.4R**) that occurred on 19th April 2007.

What can be stated summarizing all the aforementioned results presented in figures **(7 to 12)** is:

- It is made clear, that a strong correlation of the recorded signals by **PYR** and **HIO** monitoring sites, started after the 4th April 2007 and lasted up to 17th April 2007. That is thirteen **(13)** days before the occurrence of the pending earthquake.

- During this period the calculated electrical field intensity vector intersection **(IVI)** prescribed ellipses, suggesting that an electrical, preseismic signal was influencing both monitoring sites.

- This signal vanished, in accordance to the piezoelectric model, very short before the occurrence of the earthquake of 19th April 2007, M = **5.4R**, and on-wards the **(IVI)** traces turned to their normal (hyperbolic) form.

Two more examples have been analyzed in detail. The first one is the Andravida EQ, Greece (**Ms = 7.0R, 8th of June, 2008**) and the second is the Methoni EQ, Greece (**Ms = 6.0R, 21st of June, 2008**).

### 3. The data.

For both earthquakes, the normalized and band-pass filtered raw data are presented in the following figure **(13)**. The recording spans from 29th of April to 30th of June, 2008.

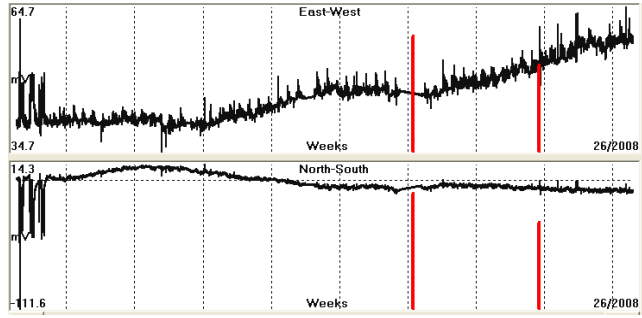

**PYR** normalized raw data

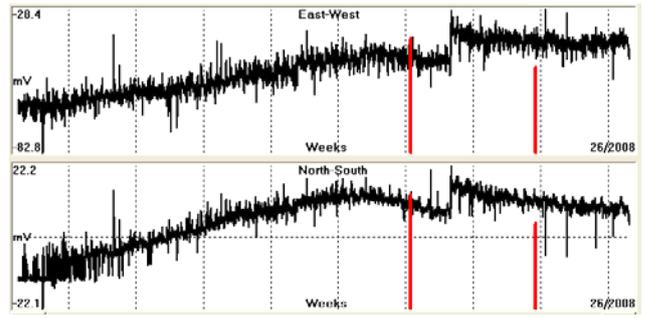

**HIO** normalized raw data

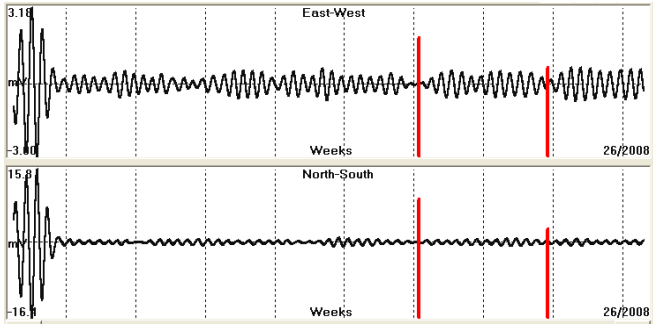

**PYR** pass-band filtered data

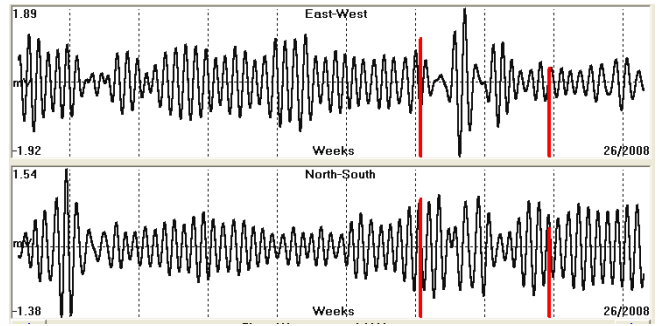

**HIO** pass-band filtered data

Fig. 13. The upper graphs represent the normalized raw data while the lower ones represent the oscillating electric field for both of Pyrgos **(PYR)** and Hios **(HIO)** monitoring sites**.** Red bars indicate the time of occurrence of both EQs.

The analysis has been made in consecutive windows of three **(3)** days. Some windows with similar results have been omitted in order to avoid making this work very lengthy. In the next figures to follow the annotation is as follows: Left diagrams = the three days period of time oscillating fields (**T=24h**) for both monitoring sites. Right figure = the corresponding phase map. For each group of drawings the corresponding recording period of time is stated. The concentric red circles indicate the location of the under consideration corresponding EQ.

**3.1. Andravida EQ (Ms = 7.0R, 8th of June, 2008).**

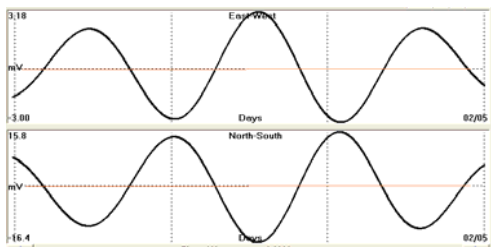

**Fig. 14. PYR 080429-080501**

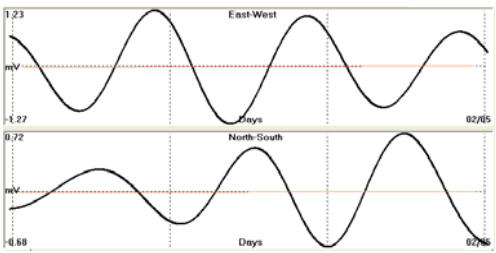

**Fig. 14a. HIO 080429-080501**

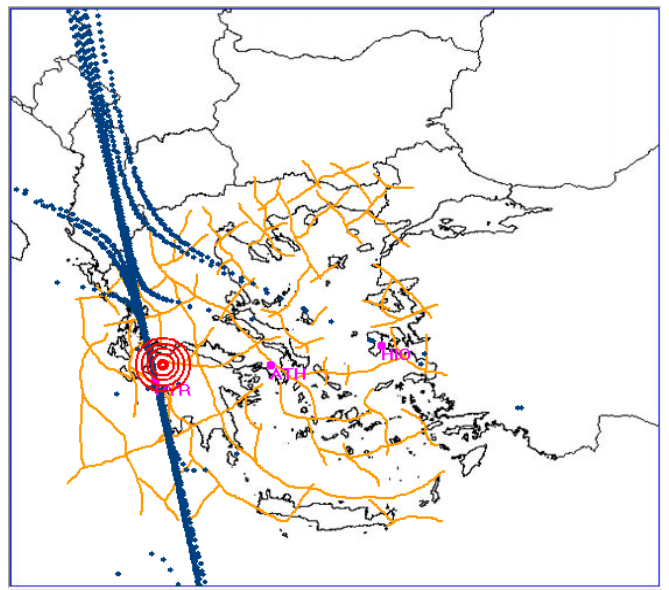

**Fig. 14b. MAP 080429-080501**

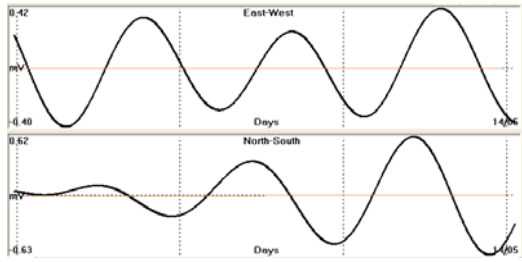

Fig. 15.  PYR 080511-080513

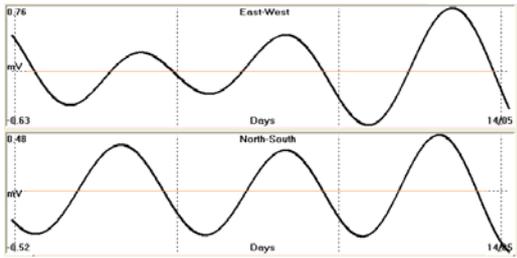

Fig. 15a. HIO 080511-080513

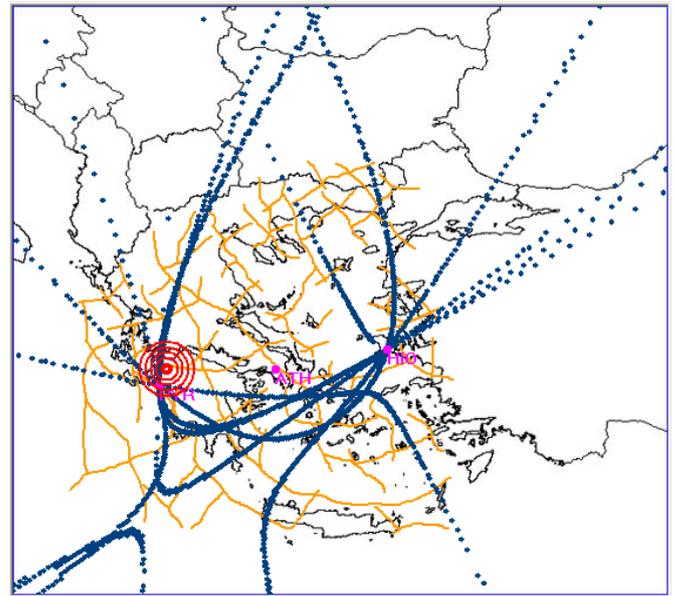

Fig. 15b. MAP 080511-080513

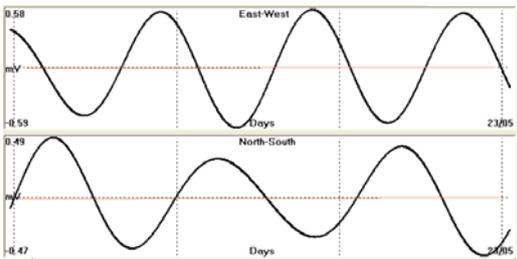

Fig. 16. PYR 080520 – 080522

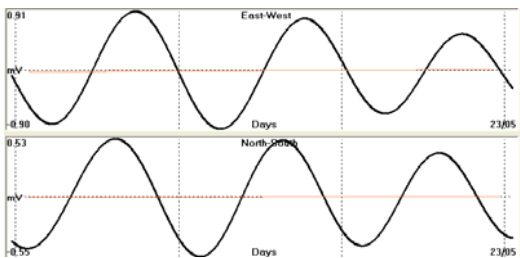

Fig. 16a. HIO 080520 – 080522

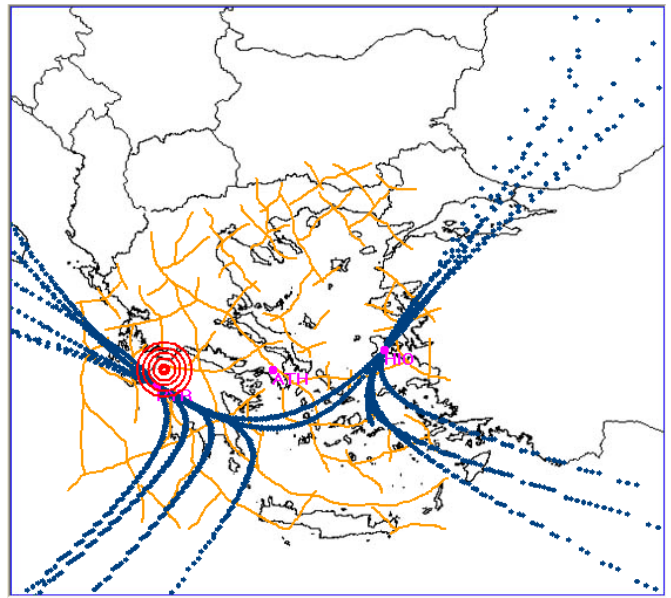

Fig. 16b. MAP 080520 – 080522

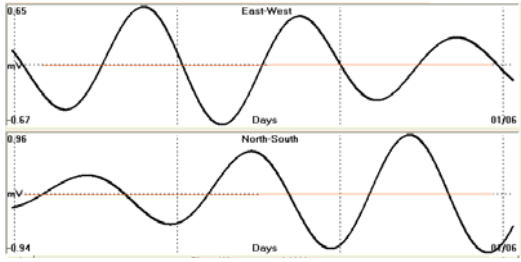

**Fig. 17. PYR 080529 – 080531**

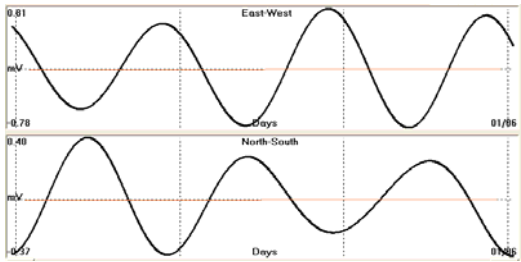

**Fig. 17a. HIO 080529 – 080531**

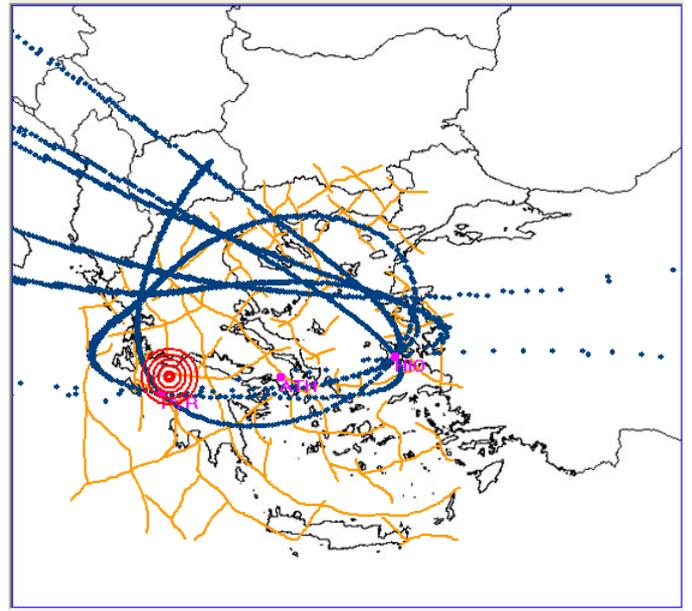

**Fig. 17b. MAP 080529 – 080531**

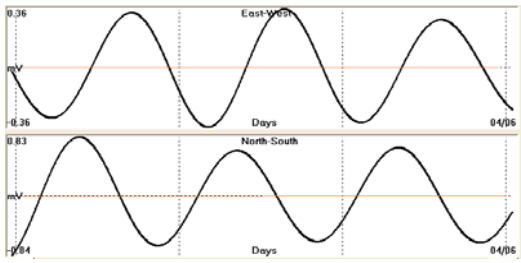

**Fig. 18. PYR 080601 – 080603**

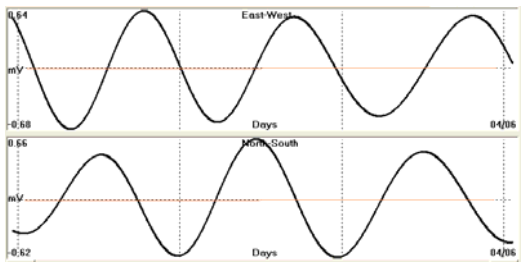

**Fig. 18a. HIO 080601 – 080603**

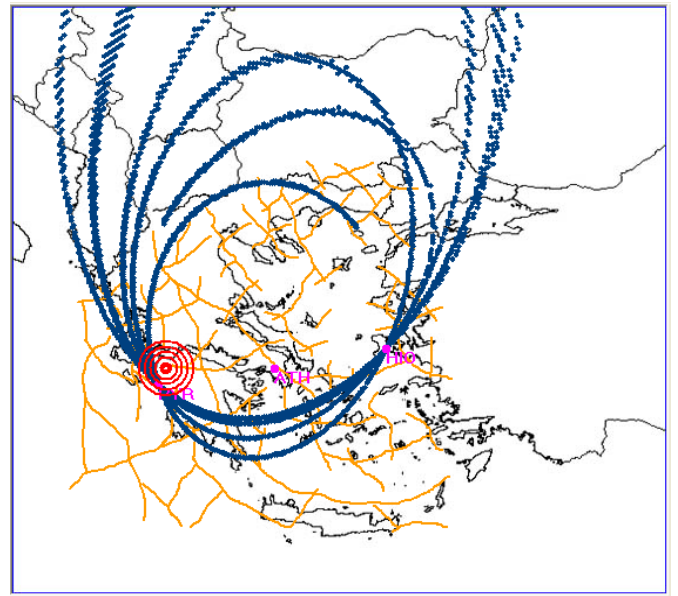

**Fig. 18b. MAP 080601 – 080603**

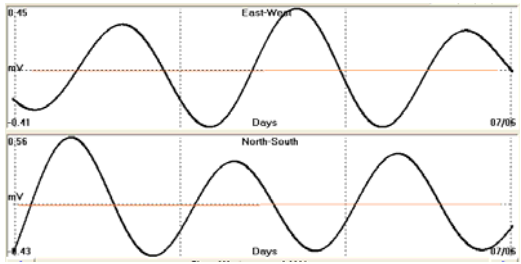

**Fig. 19. PYR 080604 – 080606**

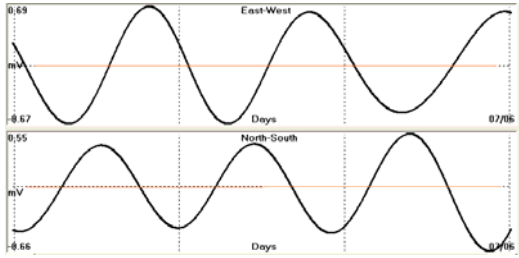

**Fig. 19a. HIO 080604 – 080606**

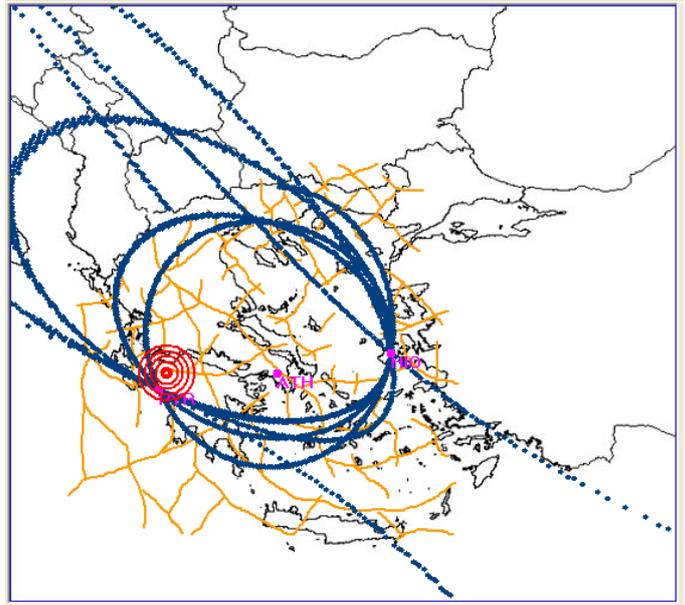

**Fig. 19b. MAP 080604 – 080606**

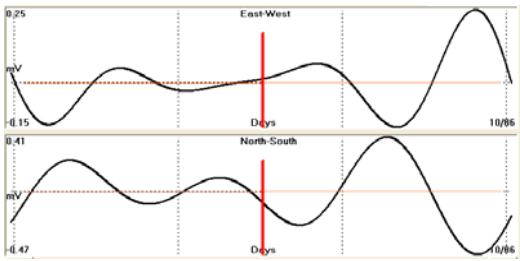

**Fig. 20. PYR 080607 – 080609**

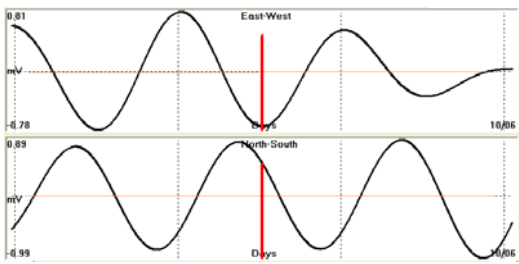

**Fig. 20a. HIO 080607 – 080609**

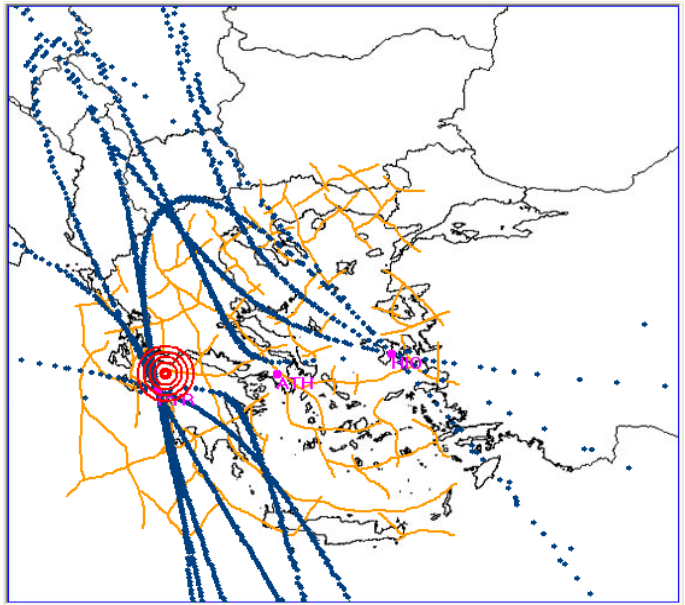

**Fig. 20b. MAP 080607 – 080609**

It is interesting to note that the EQ (red bar in fig. **20, 20a**) took place a couple of days after the ellipses vanished and the hyperbolas reappeared following the postulated methodology.

**3.2. Methoni EQ (Ms = 6.0R, 21st of June, 2008).**

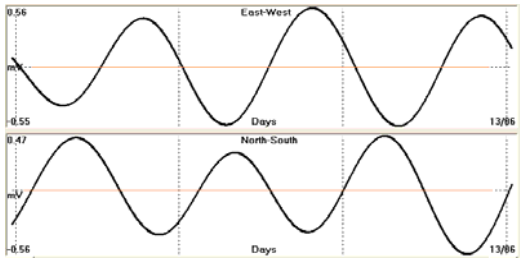

**Fig. 21. PYR 080610 – 080612**

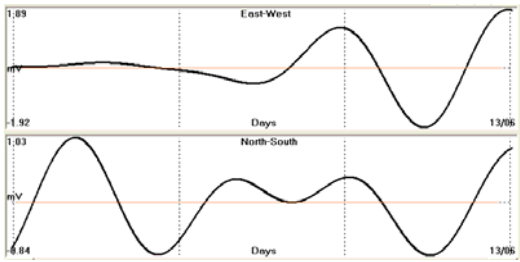

**Fig. 21a. HIO 080610 – 080612**

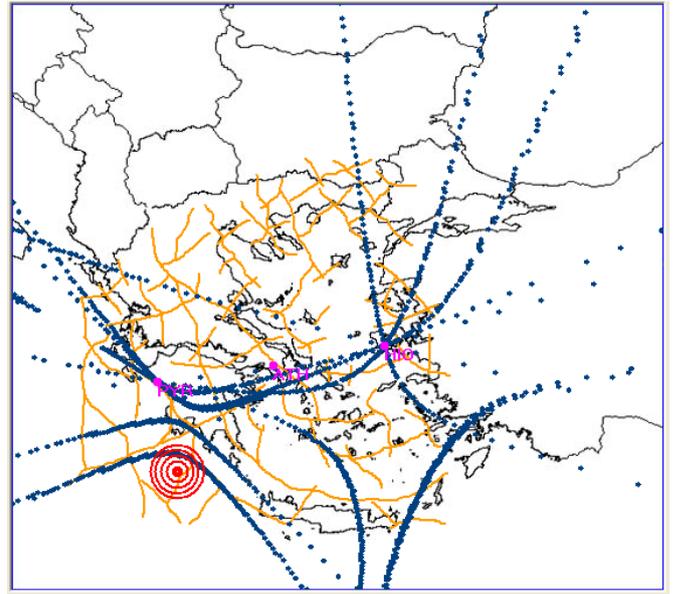

**Fig. 21b. MAP 080610 – 080612**

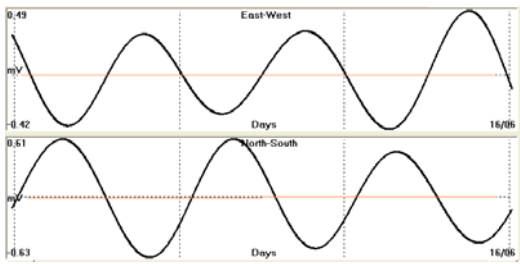

**Fig. 22. PYR 080613 – 080615**

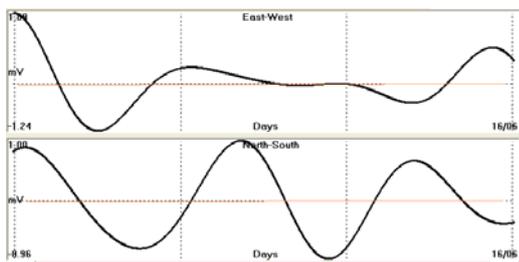

**Fig. 22a. HIO 080613 – 080615**

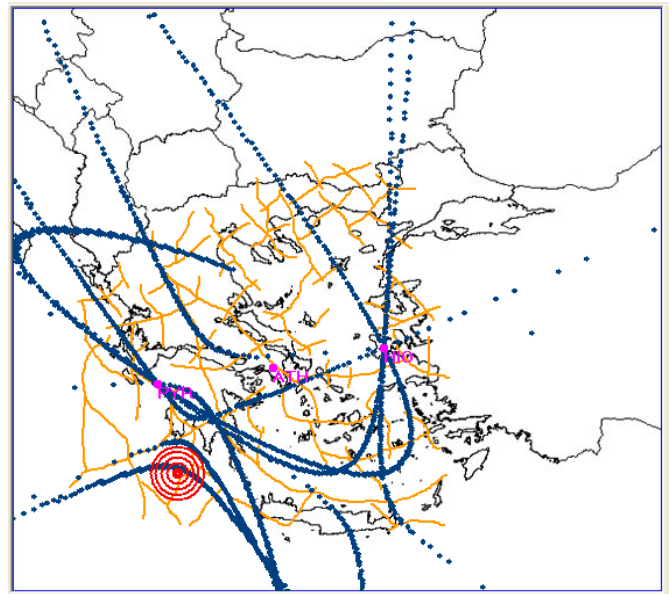

**Fig. 22b. MAP 080613 - 080615**

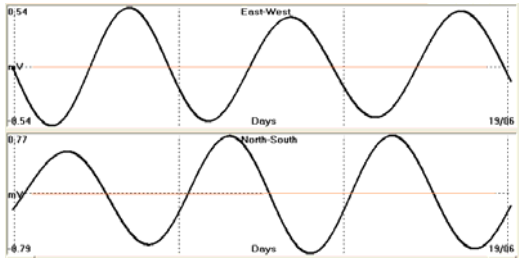

**Fig. 23. PYR 080616 – 080618**

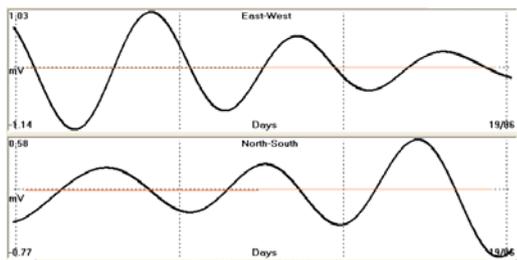

**Fig. 23a. HIO 080616 – 080618**

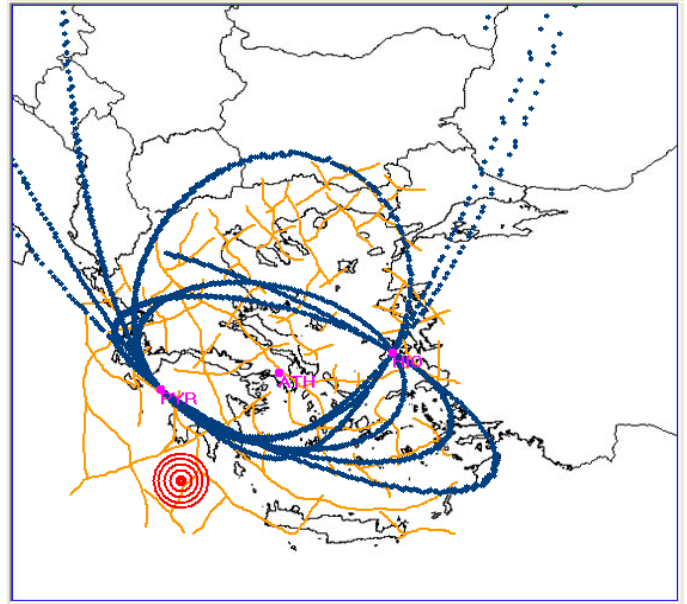

**Fig. 23b. MAP 080616 – 080618**

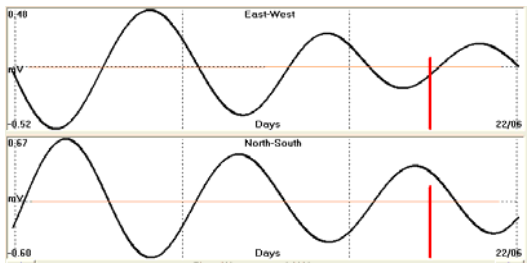

**Fig. 24. PYR 080619 – 080621**

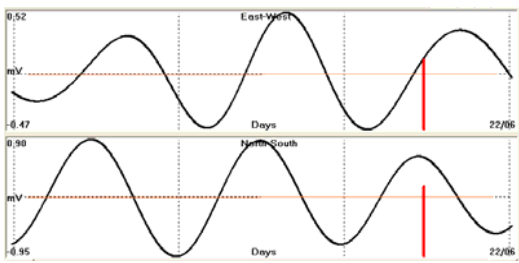

**Fig. 24a. HIO 080619 – 080621**

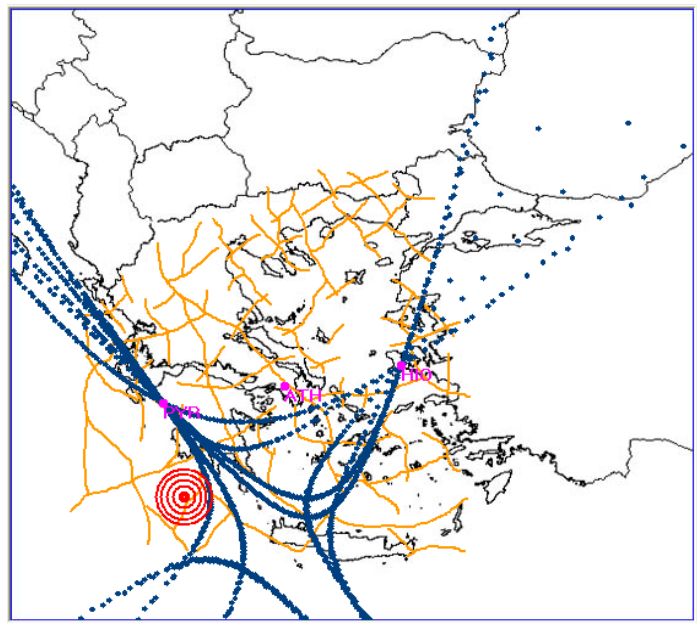

**Fig. 24b.  MAP 080619 – 080621**

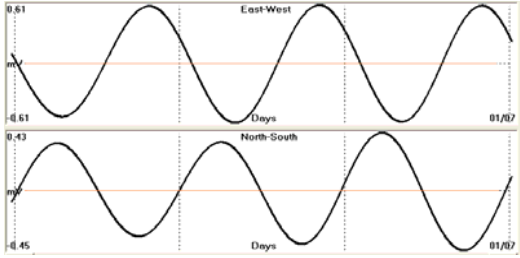

**Fig. 25. PYR 080628 – 080630**

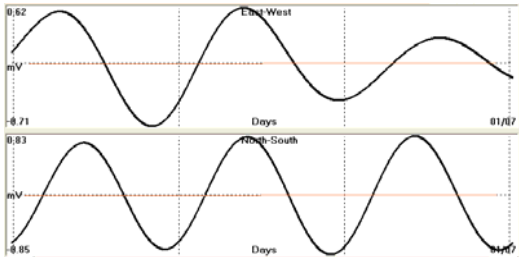

**Fig. 25a. HIO 080628 – 080630**

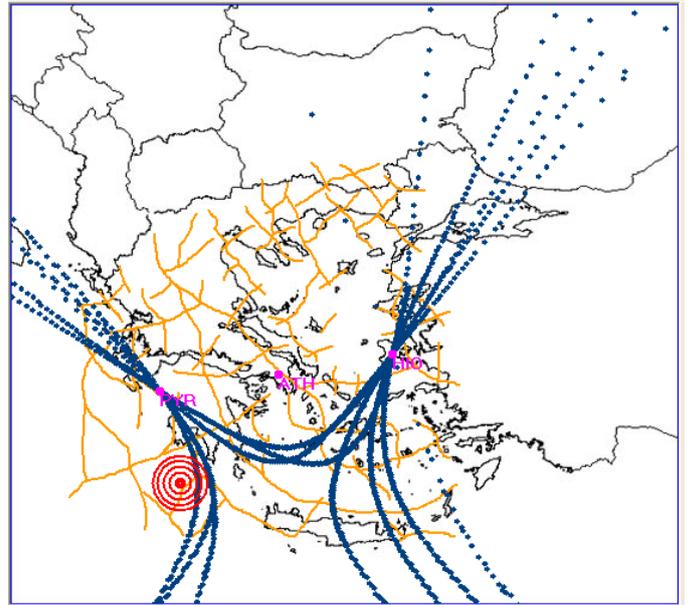

**Fig. 25b. MAP 080628 – 080630**

In this example too, the EQ (red bar in fig. **24, 24a**) took place a couple of days after the ellipses vanished and the hyperbolas reappeared following the postulated methodology.

### 4. Discussions – conclusions.

A typical problem faced by any study of recorded preseismic electric signals is the following: are the recorded signals of industrial - anthropogenic origin and therefore represent just noise or they are preseismic signals generated by any valid physical mechanism triggered at the focal area? Moreover, when the preseismic signals are of low amplitude could be easily masked by any noise present in the recording area. Additionally, if we take into account the directional properties of the electric field intensity vector (Thanassoulas, 2007), then non-preseismic electric signals will exhibit a random azimuthal direction in time, not correlated to the same time seismicity of the regional seismogenic area. On the other hand, low-level preseismic electric signals could be masked by noise so that they behave as non-preseismic ones.

In this work instead of studying one electric field time series recorded at one monitoring site, two of them, recorded at a quite large distance between each other, are correlated by techniques used for the study of non-linear dynamic systems and chaos. It was shown that the basic piezoelectric model activated in the seismogenic zone of the lithosphere justifies the generation of two different kinds of phase maps. The first one corresponds to the uncorrelated electric signals. This is characterized by pairs of mirror hyperbolas. The second type of phase map is characterized by consecutive in time ellipses. Both maps are generated by monochromatic electric field obtained after band-pass filtering of the raw normalized registered data.

The adopted physical model for this methodology, the piezoelectric one, suggests the following sequence in phase map appearance:

- Non-preseismic valid signals are present : phase map shows pairs of hyperbolas of random orientation.
  Explanation : no large EQ is expected.

- Preseismic valid signals are present : phase map shows consecutive in time ellipses.
  Explanation : A focal area has been activated thus generating preseismic electric signals.

- Preseismic valid signals vanish : phase map shows pairs of hyperbolas of random orientation.
  Explanation : The pending EQ will occur in a short time period.

The previous phases were validated by the analysis of the data obtained by **PYR** and **HIO** monitoring sites, by using the postulated methodology. The method was applied on three different large EQs that occurred in Greece on 2007 - 2008. In all three cases the above phases were clearly observed.

The study of the presented results in figures **(7)** to **(25b)** reveals some remarkable features. These are summarized in the following **Table - 1**. In this table are tabulated: **(a)** the elapsed time (in days) from the initiation of the signal till the EQ occurrence, **(b)** the duration of the signal in days, and **(c)** the time elapsed (in days) after the vanishing of the signal till the time of occurrence of the EQ.

**TABLE – 1**

| Earthquake (date, magnitude) | Signal initiation (a) | Signal duration (b) | Signal end – EQ occurrence time (c) |
|---|---|---|---|
| 19$^{th}$ April, 2007, Ms = 5.4R | 15 | 13 | 2 |
| 8$^{th}$ June, 2008, Ms = 7.0R | 11 | 9 | 2 |
| 21$^{st}$ June, 2008, Ms = 6.0R | 8 | 5 | 3 |

Although the number of EQs is very limited, for any statistical treatment of these results, it is very interesting the fact that the "strange attractor like" signals appeared well before the time of occurrence of the pending EQ. In all three cases this time varies from **(8)** days to **(15)**.

Moreover, this signal was present for a quite large period of time. In all three cases it lasted from **(5)** to **(13)** days. Finally, this signal vanished in all three cases within a very short period of time **(2 – 3 days)** before the occurrence of the pending earthquake.

Very probably, the extent of all these time periods is related to the time duration of each deformation phase of the piezoelectric model (which is presented in figure **6a**), been activated during the drastic deformation phase of the focal area.

In conclusion, this method represents a procedure that may be capable to overcome the problems met in single electric field data series, such as: masking of low level signals by noise, industrial-anthropogenic interference or any other type of noise that may exist in the monitoring sites. Thus, it enables low-level precursory electric signals to show-up much more clearly.

In terms of predictive parameters it tackles the problem of predicting the time of occurrence of a large EQ within a very short time window of a few days. A topic that is interesting to be analyzed, in more details with more examples in the future, is the "elapsed time" between vanishing of the ellipses and the time of occurrence of the pending EQ. Another interesting topic is the "ellipses duration" in respect to the magnitude or the depth of occurrence of the future large EQ. Moreover, this analysis could be combined to the observed tidal lithospheric oscillations and the generated signals with periods much larger than **(1)** day. The existence of recordings that last for very long periods of time (for some years) is an advantage to this end.

Concluding this work we can say that the use of this methodology can contribute a lot to the utilization of the determination of the occurrence time of a large EQ in terms of "short-term prediction".

## 5. References.